\documentstyle[12pt]{article}

\catcode`\@=11

\@addtoreset{equation}{section}
\def\ksection{\arabic{section}}

\def\@normalsize{\@setsize\normalsize{15pt}
\xiipt\@xiipt
\abovedisplayskip 14pt plus3pt minus3pt%
\belowdisplayskip \abovedisplayskip
\abovedisplayshortskip  \z@ plus3pt%
\belowdisplayshortskip  7pt plus3.5pt minus0pt}

\def\small{\@setsize\small{13.6pt}\xipt\@xipt
\abovedisplayskip 16pt plus3pt minus3pt%
\belowdisplayskip \abovedisplayskip
\abovedisplayshortskip  \z@ plus3pt%
\belowdisplayshortskip  7pt plus3.5pt minus0pt
\def\@listi{\parsep 4.5pt plus 2pt minus 1pt
            \itemsep \parsep
            \topsep 9pt plus 3pt minus 3pt}}

\def\underline#1{\relax\ifmmode\@@underline#1\else
	$\@@underline{\hbox{#1}}$\relax\fi}
\@twosidetrue

\catcode`@=12

\catcode`\@=11

\def\thesection{\Roman{section}}

\def\FERMIPUB{}
\def\FERMILABPub#1{\def\FERMIPUB{#1}}
\def\ps@headings{\def\@oddfoot{}\def\@evenfoot{}
\def\@oddhead{\hbox{}\hfill
	\makebox[.5\textwidth]{\raggedright\ignorespaces --
 \thepage{}--
	\hfill {\rm FERMILAB--Pub--\FERMIPUB}}}
\def\@evenhead{\@oddhead}
\def\subsectionmark##1{\markboth{##1}{}}
}

\ps@headings

\catcode`\@=12

\relax

\newcounter{appendix}

\def\appendix{\par
 \addtocounter{appendix}{1}
 \def\thesection{\Alph{appendix}}
 \def\ksection{\Alph{appendix}}}

\newcommand{\nwc}{\newcommand}
\nwc{\hyp} {\hyphenation}
\nwc{\be}  {\begin{equation}}
\nwc{\ee}  {\end{equation}}
\nwc{\ba}  {\begin{array}}
\nwc{\ea}  {\end{array}}
\nwc{\bdm} {\begin{displaymath}}
\nwc{\edm} {\end{displaymath}}
\nwc{\bea} {\be\ba{lcl}}
\nwc{\eea} {\ea\ee}
\nwc{\bda} {\bdm\ba{lcl}}
\nwc{\eda} {\ea\edm}
\nwc{\bc}  {\begin{center}}
\nwc{\ec}  {\end{center}}
\nwc{\ds}  {\displaystyle}
\nwc{\bmat}{\left(\ba}
\nwc{\emat}{\ea\right)}
\nwc{\non} {\nonumber}
\nwc{\bib} {\bibitem}
\nwc{\lra} {\longrightarrow}
\nwc{\ra}  {\rightarrow}
\nwc{\Ra}  {\Rightarrow}
\nwc{\lmt} {\longmapsto}
\nwc{\prl} {\partial}
\nwc{\iy}  {\infty}
\nwc{\ol}  {\overline}
\nwc{\hm}  {\hspace{3mm}}
\nwc{\lf}  {\left}
\nwc{\ri}  {\right}
\nwc{\lm}  {\limits}
\nwc{\lb}  {\lbrack}
\nwc{\rb}  {\rbrack}
\nwc{\ov}  {\over}
\nwc{\pr}  {\prime}
\nwc{\nnn} {\nonumber \vspace{.2cm} \\ }
\nwc{\Sc}  {{\cal S}}
\nwc{\Lc}  {{\cal L}}
\nwc{\Rc}  {{\cal R}}
\nwc{\Dc}  {{\cal D}}
\nwc{\Oc}  {{\cal O}}
\nwc{\Cc}  {{\cal C}}
\nwc{\Pc}  {{\cal P}}
\nwc{\Mc}  {{\cal M}}
\nwc{\Ec}  {{\cal E}}
\nwc{\Fc}  {{\cal F}}
\nwc{\Hc}  {{\cal H}}
\nwc{\Kc}  {{\cal K}}
\nwc{\Xc}  {{\cal X}}
\nwc{\Gc}  {{\cal G}}
\nwc{\Zc}  {{\cal Z}}
\nwc{\Nc}  {{\cal N}}
\nwc{\fca} {{\cal f}}
\nwc{\xc}  {{\cal x}}
\nwc{\Ac}  {{\cal A}}
\nwc{\Bc}  {{\cal B}}
\nwc{\Uc}  {{\cal U}}
\nwc{\Vc}  {{\cal V}}
\nwc{\Th} {\Theta}
\nwc{\th} {\theta}
\nwc{\vth} {\vartheta}
\nwc{\eps}{\epsilon}
\nwc{\si} {\sigma}
\nwc{\Gm} {\Gamma}
\nwc{\gm} {\gamma}
\nwc{\bt} {\beta}
\nwc{\La} {\Lambda}
\nwc{\la} {\lambda}
\nwc{\om} {\omega}
\nwc{\Om} {\Omega}
\nwc{\dt} {\delta}
\nwc{\Si} {\Sigma}
\nwc{\Dt} {\Delta}
\nwc{\al} {\alpha}
\nwc{\vph}{\varphi}
\def\tr{\mathop{\rm tr}}

\def\abs#1{\left| #1\right|}
\def\pr#1{#1^\prime}

\nwc{\Id}  {{\bf 1}}
\nwc{\diag} {{\rm diag}}
\nwc{\inv}  {{\rm inv}}
\nwc{\mod}  {{\rm mod}}
\nwc{\hal} {\frac{1}{2}}
\nwc{\tpi}  {2\pi i}

\def\slash#1{#1\!\!\!/\!\,\,}
\def\ap#1{Annals of Physics {\bf #1}}

\def\npb#1{Nucl. Phys. {\bf B#1}}
\def\pha#1{Physica {\bf A#1}}
\def\plb#1{Phys. Lett. {\bf B#1}}
\def\prd#1{Phys. Rev. {\bf D#1 }}
\def\prle#1{Phys. Rev. Lett. {\bf #1}}
\def\ptp#1{Progr. Theor. Phys. {\bf #1}}
\def\rmp#1{Rev. Mod. Phys. {\bf #1}}
\def\zpc#1{Z. Phys. {\bf C#1}}

\def\GeV {\,{\rm  GeV}}

\def \lta {\mathrel{\vcenter
     {\hbox{$<$}\nointerlineskip\hbox{$\sim$}}}}
\def \gta {\mathrel{\vcenter
     {\hbox{$>$}\nointerlineskip\hbox{$\sim$}}}}
\newsavebox{\nnin} \sbox{\nnin}{$\hspace{1mm}
\in\kern -.8em /
                   \hspace{1mm}$}

\newcommand{\sub}{\subset}
\newsavebox{\nnsub} \sbox{\nnsub}{$\hspace{1mm}
            \sub\kern -.9em /
            \hspace{1mm}$}

\def\KK{{\rm I\kern -.2em  K}}
\def\NN{{\rm I\kern -.16em N}}
\def\RR{{\rm I\kern -.2em  R}}
\def\ZZ{Z \kern -.43em Z}
\def\QQ{{\rm \kern .25em
             \vrule height1.4ex depth-.12ex width.06em
             \kern-.31em Q}}
\def\CC{{\rm \kern .25em
             \vrule height1.4ex depth-.12ex width.06em
             \kern-.31em C}}
\def\ZZZ{Z\kern -0.31em Z}

\begin{document}
\par \vskip .05in
\FERMILABPub{93/253--T}
\begin{titlepage}
\begin{flushright}
FERMI--PUB--93/253--T\\
{\today}
\end{flushright}
\vfill
\begin{center}
{\large \bf Inhomogeneous Field Configurations \\
and the \\
Electroweak Phase Transition}
 \end{center}
  \par \vskip .1in \noindent
\begin{center}
{\bf Dirk--U. Jungnickel\footnote{Supported
  by the Deutsche Forschungsgemeinschaft}}
  \par \vskip .05in \noindent
{Fermi National Accelerator Laboratory\\
 P.O. Box 500, Batavia, Illinois, 60510}
  \par \vskip .05in \noindent
 {\bf Dirk Walliser\footnote{Supported
  by Wilhelm--Zangen--Fonds} }
  \par \vskip .05in \noindent
{NASA/Fermilab Astrophysics Center\\
 Fermi National Accelerator Laboratory\\
 P.O. Box 500, Batavia, Illinois, 60510}
  \par \vskip .05in \noindent
\end{center}
\begin{center}{\large Abstract}\end{center}
\par \vskip .05in
\begin{quote}
We investigate the effects of inhomogeneous
scalar field
configurations on the electroweak phase
transition. For this
purpose we calculate the leading
perturbative correction to the
wave function correction term $Z(\vph,T)$,
i.e., the kinetic term in the effective action,
for the electroweak
Standard Model at finite temperature
and the top quark self--mass.
Our finding for the fermionic contribution
to $Z(\vph,T)$ is infra--red finite and disagrees
with other recent results.
In general, neither the order
of the phase transition nor the temperature at which it occurs
change, once $Z(\vph,T)$ is included. But a
non--vanishing,
positive (negative) $Z(\vph,T)$
enhances (decreases) the critical
droplet surface tension and the strength
of the phase transition.
We find that in the range of parameter space,
which allows for a
first--order phase transition, the wave
function correction term is
negative --- indicating a weaker phase
transition --- and
especially for small field values so
large that perturbation theory
becomes unreliable.
\end{quote} \vfill \end{titlepage}

\section{Introduction}

The understanding of the electroweak phase
transition has matured
rapidly during recent years. The original work
on restoration of
gauge symmetries at high temperatures \cite{KL}
and systematic,
partial summation of perturbation theory
\cite{Weinberg} has
attracted a lot of attention in the
framework of the Standard
Model, since the suggestion of
a mechanism for baryo--genesis at
the electroweak scale \cite{KRS}.
Already in the early
eighties it became clear that
non--abelian gauge theories contain
massless degrees of freedom in
perturbation theory at high
temperatures \cite{GPY}.
Therefore, infrared problems complicate
the straightforward setup of
perturbation theory and a considerable
effort was made recently to resolve them
\cite{EIKR}--\cite{EQZ}.
Up to Higgs masses of about
$80\GeV$, these difficulties
may be cured by introducing improved
propagators (c.f. e.g.,
\cite{BHW}--\cite{EQZ}). In the
improved perturbation theory
self--energy and self--mass
corrections are included in
the action from the beginning. They are
determined by solving appropriate
special cases of the
Dyson--Schwinger equations, the gap equations
\cite{Weinberg,BFHW,EQZ}.
Keeping so--called ``hard thermal loops''
and summing over ``soft thermal
loops'' the leading infrared
singularities can be shown to
cancel out \cite{Pisarski}.

The order of the phase transition
depends crucially on the value of
the magnetic mass of the  gauge
bosons, whose calculation requires
non--perturbative techniques.
Although different approaches suggest
the same order of magnitude
\cite{BFHW,BLS}, the determination of
its value to within more then
$10\%$ accuracy would be desirable.
As long as the magnetic mass
is smaller than $0.07\ T$, where
$T$ denotes the temperature of
the heat bath, the phase transition
appears to be of first order
in the range of Higgs masses
accessible to perturbation theory.
However, it is still an open
question, whether the phase
transition within the electroweak
standard model is strong enough
to account for the observed baryon
asymmetry of the universe \cite{FS}.
Due to its weakness the phase
transition may even proceed via
formation of sub--critical droplets
\cite{GKW}.

All the above results have been
extracted purely from the effective
potential $V_{\rm eff}(\vph,T)$,
i.e., the effective action for
vanishing derivative terms
$\prl_\mu\vph=0$ of the scalar field
$\vph$, which plays the role
of the order parameter.
In this paper we investigate
whether the full effective action is
indeed dominated by homogeneous
field configurations and whether it
is justified to neglect quantum
corrections which give rise to
derivative terms in the effective
action. We wish, however, to
point out from the beginning that,
as a matter of fact, a
consistent one--loop treatment of
the phase transition requires the
inclusion of the wave function
correction term whose leading
contribution occurs at this
level in perturbation theory.
Indeed, it was argued in refs.
\cite{DG}--\cite{KRS2} that
depending on the theory under
consideration the expansion of the
effective action around $\prl_\mu\vph=0$
might break down and
non--perturbative effects
become important. In addition we will see
that quantities essential for
the mechanism of critical droplet
nucleation such as the droplet
surface tension and the strength of
the phase transition are modified
by the presence of higher
derivative terms. Although,
in this paper, we will perform a
perturbative calculation there
are techniques of averaging the
action over a range of momenta
without expanding it in terms of
derivatives \cite{TW}.

The outline of the paper is as follows:
In section \ref{Inhom} we will
introduce some essential quantities
for the description of
first--order phase transitions, review their
derivation for homogeneous
scalar field configurations and extend
the analysis to inhomogeneous
field configurations.
In particular, we explain the
impact of the wave function
correction factor $Z(\vph,T)$ on
droplet nucleation rate and surface
tension for a first--order
phase transition. In section
\ref{SelfEnergyCorrections} we
discuss the self--energy corrections for
the various particles in the theory
in the presence of a plasma
which help to cut off infrared
divergences. A general method to
derive the kinetic term in the
effective action developed in \cite{MTW}
is reviewed in section \ref{LocMomentExp}
where we explicitly calculate the
different contributions to the wave
function correction term for the Standard Model.
Our results are described in
section \ref{Results}, and our
conclusions are presented in section
\ref{Conclusions}.
A discussion
of the top quark self--mass
and several useful integrals with their
high--temperature expansions
have been relegated to two appendices.

\section{The wave function correction term}
\label{Inhom}

\subsection{Decay of metastable states}
\label{DecayMetaStates}

The entire dynamics of the phase
transition is contained in the
quantity
\be
 Z_\beta [\vph]=\int[\Dc\hat{\vph}]
 [\Dc W_\mu][\Dc\psi]
 \exp\left\{ S_\beta [\vph+\hat{\vph},
 W_\mu,\psi]\right\} \; .
 \label{PartFunct0}
\ee
The path
integral is performed over
fluctuations $\hat{\vph}$ around a
classical field configuration
$\vph(x)$, over vector and ghost
fields, whose measure is collectively
denoted by $[\Dc W_\mu]$ and
over fermion fields with measure $[\Dc\psi]$.
The exponent contains
the classical  action
$S_\beta$ at finite temperature $T=\beta^{-1}$. An
additional term that vanishes for stationary field
configurations has been neglected.

Following the approach of \cite{BFHW},
in a first step we integrate
out all vector, ghost and fermion
fields (but not the scalar
field fluctuation $\hat{\vph}$)
to arrive at an effective
(coarse--grained) finite--temperature
action $\Gm_\beta [\vph]$
defined via
\be
 Z_\beta [\vph]=:\int[\Dc\hat{\vph}]
 \exp\left\{\Gm_\beta [\vph+\hat{\vph}]\right\}\; ,
 \label{PartFunct}
\ee
which may be expanded in
powers of derivatives of $\vph$:
\be
 \Gm_\beta [\vph]=\int_\beta\left\{\hal (\prl\vph)^2
 -V(\vph,T)\right\}
 +\int_\beta\sum_{n=2}^\infty
 \frac{1}{n!}Z_n(\vph,T)(\prl\vph)^n \; .
 \label{DerExpansion}
\ee
Each $Z_n$ may in addition be
expanded in a power series in the
coupling constants of the theory.

The
remaining integration over
$\hat{\vph}$ in (\ref{PartFunct}) will
be carried out in the saddle
point approximation. We collectively
denote all possible Lorentz
invariant derivative terms of
$\vph$ with $n$
derivatives\footnote{For simplicity we have not
taken into account the fact
that for given $n$ different combinations of
derivatives may
have different prefactors $Z_n$.} by
$(\prl\vph)^n$ and use the shorthand notation
\be
 \int_\beta =\int_0^\beta d\tau\int d^3 x \; .
\ee
Throughout this paper we will use
the imaginary time formalism
\cite{Matsubara,Kapusta}. Therefore
boson and fermion fields
satisfy periodic and anti--periodic
boundary conditions in the
imaginary time $\tau=ix_0$, respectively.
In momentum space this
leads to integrals
\be
 \sum_{\om}\frac{1}{\beta}
 \int\frac{d^3 k}{(2\pi)^3}
\ee
where $\om=2\pi inT$
($\om=(2n+1)\pi iT$) for bosonic (fermionic)
fields and $n$ runs over all integers.
The first term in (\ref{DerExpansion}) contains the
effective potential $V(\vph,T)$
and the classical kinetic term,
whereas the second summand
incorporates derivative terms due to
quantum corrections.

For stationary fields, $\prl_\tau\vph =0$,
the effective
action $\Gm_\beta [\vph ]$
plays the role of the free energy
\be
 F[\vph ,T]=-\frac{1}{\beta}\Gm_\beta [\vph ] \; ,
\ee
and $\vph$ is the order
parameter of the phase transition.
Usually first--order phase
transitions are studied under the
assumption that the effective
action $\Gm_\beta[\vph]$ is dominated
by homogeneous field configurations, i.e., the
derivative terms proportional
to $Z_n$ in (\ref{DerExpansion}) are
neglected. Then the remaining
path integral in (\ref{PartFunct})
is carried out in a saddle point approximation around
configurations $\ol{\vph}(\vec{x})$
which extremize the classical
free energy
\be
 F[\vph ,T]=\int d^3 x\left\{ V(\vph ,T)+\hal
 \abs{\vec{\nabla}\vph}^2\right\} \; .
 \label{ClassFreeEnergy}
\ee
Isotropic configurations
$\ol{\vph}(r\equiv\abs{\vec{x}})$
obey the differential equation
\be
 \frac{d^2 \ol{\vph}}{dr^2}+
 \frac{2}{r}\frac{d\ol{\vph}}{dr}
 -\frac{\prl V}{\prl\vph}(\ol{\vph}) =0
 \label{dropletDGL}
\ee
which is obtained by varying
(\ref{ClassFreeEnergy}) w.r.t. $\vph$.

The generic potential for a
first--order phase transition
and for $T_b<T<T_c$ has two local
minima (c.f. Fig. \ref{Veff}) one of
which is metastable. Here $T_c$
denotes the critical temperature,
where the minima are degenerate,
and $T_b$ is the so--called
barrier temperature at which the
potential barrier between the
minima vanishes. We choose the
metastable minimum to occur at $\vph
=0$ and denote the position of
the global one by $\vph =\vph_{\rm
min}(T)$. The appropriate
boundary conditions for the tunneling
solution, which  interpolates
between the symmetric ($\vph=0$) and
the broken ($\vph\neq 0$) phase,
are $\ol{\vph}^{\prime}(r=0)=0$
and $\ol{\vph}(r\ra\infty)=0$
\cite{Coleman}. In the thin wall
approximation \cite{Linde} one finds
\be
 \ol{\vph}(r)=\hal\vph_{\rm min}\left[1-
 \tanh\left(\frac{r-R(T)}{d}\right)\right]
 \label{Bounce}
\ee
where
\be
 R(T)=\frac{2\si}{\Dt V(T)}
\ee
is the droplet radius at
temperature $T$. It depends on the
surface tension $\si(T)$
which may be evaluated at the critical
temperature $T_c$
\be
 \si=\int_0^{\vph_{\rm min}(T_c)} d\vph
 \sqrt{2V(\vph,T_c)}
\ee
and the potential difference between the two minima
\be
 \Dt V(T)=V(0,T)-V(\vph_{\rm min},T)\; .
  \label{PotDiff}
\ee
The thickness $d$ of the droplet
wall depends on the detailed shape
of the potential. For a polynomial of the form
\be
 V(\vph,T)=\hal m^2(T)\vph^2
 -ET\vph^3+\frac{1}{4}\la\vph^4
 \label{ToyPotential}
\ee
one obtains
\be
 d=\sqrt{\frac{2}{\la}}\frac{2}{\vph_{\rm min}} \; .
\ee
By inserting (\ref{Bounce})
into (\ref{ClassFreeEnergy}) we find the
free energy in the thin wall
approximation as a sum of a surface
and a volume term:
\be
 F_{\rm TW}[\ol{\vph},T]
 =4\pi R^2(T)\si -\frac{4\pi}{3}R^3(T)\,\Dt V(T)
 =\frac{4\pi}{3}\si R^2(T)\; .
\ee

Once the thin wall approximation
breaks down, $\Dt V(T)$ becomes
larger than the barrier between
the minima, and (\ref{dropletDGL})
has to be solved numerically.
Unfortunately, in the Standard
Model the thin wall approximation is
only marginally applicable for $m_H\simeq 80\GeV$
\cite{DLHLL,BFHW}.

To second order in the saddle
point approximation of
(\ref{PartFunct}) we need to
perform a Gaussian path integral over
the fluctuations $\hat{\vph}$
around the tunneling solution
$\ol{\vph}$. This correction
appears as a pre--exponential factor
$A$ in the nucleation rate
of critical droplets
\be
 \Gm(T) = AT^4 \exp\left\{
 -\beta\left( F[\ol{\vph},T]-
 F[0,T]\right)\right\} \; .
 \label{BubNucRate}
\ee
It has been shown to be of order
one for the electroweak phase
transition \cite{DLHLL,BFHW}.
In the following we will
therefore only be concerned with the saddle
point configuration $\ol{\vph}(r)$.

The temperature $T_e$ which
marks the end of the phase transition
may be defined as the temperature
for which the droplet nucleation
rate becomes larger than the
expansion rate of the universe, i.e.,
$\Gm(T_e)\gta H^4(T_e)$, where
$H(T)$ denotes the Hubble function.
Using (\ref{BubNucRate}) this
leads to the rough estimate $\beta_e
F[\ol{\vph},T_e]\lta 145$.

\subsection{Inhomogeneous field configurations}

Higher order derivative corrections
will in general alter the
results of the last section.
Some recent publications
\cite{DG,EEV} indicate that depending on
the parameters of the theory
under consideration they may even
dominate the leading terms
and the expansion breaks down
altogether. But even in the
domain where the perturbative results
seem to signal that this is
not the case, non--perturbative
considerations yield results
which are different from perturbative
calculations for the
effective potential alone
\cite{Shaposhnikov,KRS2}.
In this paper we will restrict ourselves
to the first term of the
derivative expansion, i.e.,
$Z(\vph,T):=Z_2(\vph,T)$,
that corrects the scalar wave function.
Here we will merely discuss
possible physical consequences of a
non--vanishing $Z(\vph,T)$ and
leave the explicit calculations
within the Standard Model to
the forthcoming sections. We would
like to point out that a
consistent one--loop analysis of the
electroweak phase transition
indeed requires the inclusion of
$Z(\vph,T)$, since its leading
perturbative contribution occurs at
the one--loop level.

To check whether homogeneous
field configurations are important for
the dynamics of the phase transition we
wish to investigate how
$Z(\vph,T)\neq 0$ affects the
characteristic quantities
of the phase transition. For this
purpose we perform the scalar field transformation
\be
 \tilde{\vph}(r)=\int d\vph\sqrt{1+Z(\vph,T)}
 \label{ScalarTrans}
\ee
under which the free energy
\be
 F[\tilde{\vph},T]=\int d^3 x\left\{
 \tilde{V}(\tilde{\vph},T)+\hal
 (\vec{\nabla}\tilde{\vph})^2\right\}
 \label{CorrFreeEnergy}
\ee
has the same form as for $Z(\vph,T)=0$.
The new potential
$\tilde{V}(\tilde{\vph},T)=
V(\vph(\tilde{\vph}),T)$ is locally
rescaled. Since
$\frac{\prl\tilde{\vph}}{\prl\vph}=\sqrt{1+Z}>0$
there is a one--to--one
correspondence between $\tilde{\vph}$ and
$\vph$, i.e., (\ref{ScalarTrans})
amounts to a local rescaling of
the $\vph$--axis in Fig. \ref{Veff}.
Minima and maxima still have
the same potential energy
$\tilde{V}(\tilde{\vph}_{\rm
min},T)=V(\vph_{\rm min},T)$
and as a consequence the critical
temperature $T_c$, at which
the two minima are degenerate, does not
change either. Also neither
the height of the barrier nor the
amount of supercooling
$\Dt\tilde{V}(T)=\Dt V(T)$ change. Thus, once
the thin wall approximation
is valid for homogeneous field
configurations it survives
the incorporation of the wave function
correction term.

However, the new surface tension
\be
 \tilde{\si} =\int_0^{\tilde{\vph}_{\rm min}}
 d\tilde{\vph}
 \sqrt{2\tilde{V}(\tilde{\vph},T)}=
 \int_0^{\vph_{\rm min}}d\vph
 \sqrt{2\left[ 1+Z(\vph,T)\right] V(\vph,T)}
 \label{NewSurfTens}
\ee
may be substantially
different from the surface tension $\si$
without wave function correction term.
Hence, a possible measure
for the effect of the wave function
correction term on the
dynamics of the phase transition is given by
\be
 \dt_Z F:=
 \frac{F[\ol{\tilde{\vph}},T]
 -F[\ol{\vph},T]}{F[\ol{\vph},T]}
 =\frac{\tilde{\si}^3-\si^3}{\si^3}
 \label{Criterion}
\ee
where $\ol{\tilde{\vph}}$
denotes the stationary isotropic solution
which extremizes the corrected
free energy (\ref{CorrFreeEnergy}).
For a constant $Z=0.25$,
for instance, one would obtain the
significant relative deviation
of $\dt_Z F\simeq 0.4$.

Another important consequence
of a non--negligible $Z$--factor is
the end temperature of the
phase transition. Suppose that
$Z(\vph,T)>0$; this would
result in an increased surface tension
and the universe would have
to supercool further to complete the
phase transition:
$\tilde{T}_e <T_e$. Therefore, the
strength of the phase
transition would increase, since
$\frac{\tilde{\vph}_{\rm min}
(\tilde{T}_e)}{\tilde{T}_e}
>\frac{\vph_{\rm min}(T_e)}{T_e}$,
and could be more
favorable for baryogenesis
than anticipated without wave function
corrections.  A rough estimate
based on the potential
(\ref{ToyPotential}) and the
assumption of an approximately
constant $Z$ yields
\be
 \frac{\tilde{\vph}_{\rm min}
 (\tilde{T}_e)}{\tilde{T}_e}=
 \left\{ 1+\frac{1}{6}\left[\left(
 1+Z\right)^{\frac{3}{4}}-1\right]\right\}
 \frac{\vph_{\rm min}(T_e)}{T_e} \; ,
 \label{PTStrength}
\ee
where we assumed that
$\frac{\la m^2 (T_e)}{9ET_e}<<1$.
Eqn. (\ref{PTStrength})
shows that a positive wave function
correction factor $Z$ would
enhance the strength of a first--order
phase transition whereas a
negative $Z$ would decrease it.

\section{Self--energy corrections and
improved propagators}
\label{SelfEnergyCorrections}

Finite--temperature field theory
is known to develop infrared
singularities once massless
degrees of freedom are present. Since
in the symmetric phase all
masses are essentially zero,
straightforward perturbation
theory breaks down. This problem can
be avoided by using an
improved perturbation theory with
propagators that include
finite--temperature self--energy and
self--mass corrections
\cite{BFHW}--\cite{Pisarski}, \cite{Altherr}
generically denoted by
$\Si_n(x)$ in this paper. In general, the
full propagator $\Dc_n(k)$
is determined by the Dyson--Schwinger
equation
\be
 \Dc_n^{-1}(k)=\Dc_{n,0}^{-1}(k)+i\Si_n(k)\; ,
 \label{SchwingerDysonEqn}
\ee
where $n$ labels the different
fields in the theory and
$\Dc_{n,0}(k)$ is the tree--level
propagator. The Dyson--Schwinger
equations may be solved
perturbatively. To the order to
which we are calculating, the vertex
functions can be replaced by the
tree level couplings $g_i$ and we
are left with a one--loop diagram for $\Si_n$.

In the following we will
simply state the known self--energies
$\Si_n=\Pi_s(k)$ for scalars
and $\Si_n=-\Pi_v(k)$ for vectors in
the electroweak Standard Model.
The top quark self--mass
$\Si_n=\Si_t(k)$ is evaluated
in appendix \ref{topselfen} in the
limit of small $k$.

The starting point for our
explicit analysis is the simplified
$SU_L(2)$ Standard Model (SM)
with vanishing hypercharge gauge
coupling $g^\prime$. In this
approximation the $SU_L(2)$ gauge bosons
are degenerate in mass and
the Weinberg angle $\Theta_W$ is
neglected.  The corresponding Lagrangian is
\be
 \Lc =\Lc_{\rm gauge}
 +\Lc_{\rm Higgs}+\Lc_{\rm fermion}
 +\Lc_{\rm GF}+\Lc_{\rm ghost}
 \label{SMLagrangian}
\ee
with
\be
 \Lc_{\rm gauge}+\Lc_{\rm Higgs}=
 -\frac{1}{4}F_{\mu\nu}^a F^{a,\mu\nu}+\
 \left( {\rm D}_\mu \Phi\right)^\dagger
 \left( {\rm D}^\mu\Phi\right)
 -\mu^2\left(\Phi^\dagger\Phi\right)
 -\la\left(\Phi^\dagger\Phi\right)^2\; ,
\ee
$\mu^2 <0$ and
\be
 {\rm D}_\mu =\prl_\mu -ig\frac{\tau^a}{2}W_\mu^a\; ,
\ee
where the $\tau^a$, $a=1,2,3$,
denote the three Pauli matrices.
The field $\Phi$ is an $SU_L(2)$
doublet parameterized by
four real scalar fields:
\be
 \Phi (x)=\frac{1}{\sqrt{2}}
 \left(\ba{c}
 \chi_1(x) +i\chi_2(x)\\ \vph (x)+h(x)+i\chi_3(x)
 \ea\right)\; ,
\ee
where $h$ is the Higgs field,
$\chi_a$, $a=1,2,3$, are the
three Goldstone bosons and
$\varphi$ is a real background
field.

In the fermionic part of
the Lagrangian we neglect
all lepton and quark Yukawa
couplings compared to the
top quark Yukawa coupling $f_t$. Hence
\be
 \Lc_{\rm fermion}=\ol{\psi}_L
 i{\rm D}_\mu\gm^\mu\psi_L
 +f_t\left(\ol{t}_L, \ol{b}_L\right)
 \left( i\tau_2\Phi^*\right) t_R +{\rm h.c.}
\ee
The gauge fixing and
ghost Lagrangians are given by
\bea
 \Lc_{\rm GF} &=& \ds{
 -\frac{1}{2\xi}\Fc_a \Fc^a}\nnn
 \Lc_{\rm ghost} &=& \ds{\ol{c}_a\Mc^{ab}c_b}
\eea
with $\Fc_a(W)$ and $\Mc$ defined via
\bea
 \Mc^{ab} &=& \ds{\left. \frac{\dt}{\dt\om_a}
 \Fc^b (W^\om)\right|_{\om =0}}\nnn
 \Fc_a &=& \ds{\prl_\mu W_a^\mu
 -\hal g\xi\varphi\chi_a}
\eea
where $W^\om$ denotes the result
of an infinitesimal gauge
transformation $U(\om )=1-iT_a\om^a$
on the gauge field $W$.
Throughout this paper we will
work in Landau gauge $\xi =0$.

To one--loop the full
scalar propagator can easily be read off from
(\ref{SchwingerDysonEqn}):
\be
 \Dc_{\vph,\chi}(k)=
 \frac{i}{k^2-m_{\vph,\chi}^{(0)2}
 -\Pi_{\vph,\chi}(k)}
 \label{ImprovedScalarProp}
\ee
with $m_\vph^{(0)2}=\la (3\vph^2-v^2)$,
$m_\chi^{(0)2}=\la
(\vph^2-v^2)$ being the
tree--level scalar masses. Due to the
breakdown of Lorentz invariance
in the presence of a heat bath the
full vector propagator ($m_W^{(0)}=g\vph /2$)
\bea
 \Dc^{\mu\nu}(k) &=& \ds{\frac{-i}
 {k^2-m_W^{(0)2}-\Pi_L(k)}P_L^{\mu\nu}}\nnn
 &+& \ds{\frac{-i}{k^2-m_W^{(0)2}
 -\Pi_T(k)}P_T^{\mu\nu}
 +\frac{-i\xi}{k^2-\xi m_W^{(0)2}
 -\Pi_G(k)}P_G^{\mu\nu}}
 \label{GaugeProp}
\eea
involves projections onto
the direction $k_\mu$ of propagation
($P_G^{\mu\nu}$), onto the
component of the heat bath flow $u_\mu$
perpendicular to $k_\mu$
($P_L^{\mu\nu}$) and onto the remaining
two directions ($P_T^{\mu\nu}$).
For explicit definitions and
properties c.f. \cite{BFHW}.
In terms of these projectors the
self--energy tensor
\be
 \Pi^{\mu\nu}(k)=\Pi_L(k)P_L^{\mu\nu}
 +\Pi_T(k)P_T^{\mu\nu}+\Pi_G(k)P_G^{\mu\nu}
 +\Pi_S(k)S^{\mu\nu}
 \label{WSelfEn}
\ee
also involves a traceless
projector $S$ which does not contribute
to $\Dc_{\mu\nu}$ in Landau
gauge. Note that we keep the third
term in (\ref{GaugeProp})
although at first sight it seems to
vanish for our gauge choice
$\xi\ra 0$. Eventually one encounters
singularities in the gauge
fixing contribution of $Z(\vph,T)$ for
$\xi\ra 0$ which are precisely
canceled by this term.

In the limit $k\ra0$ the
self--energies give rise to the plasma
mass corrections
\bea
 \Pi_{\vph,\chi}(0)&=&\ds{\dt m_{\vph,\chi}^2
 =\left(\frac{3}{16}g^2+\hal\la
 +\frac{1}{4}f_t^2\right)T^2}\nnn
 \Pi_L(0)&=& \ds{\dt m_{W,L}^2
 = \frac{11}{6}g^2 T^2}\nnn
 \Pi_T(0)&=& \ds{\dt m_{W,T}^2
 =\frac{1}{9\pi^2}\gm^2g^4T^2}\; .
\eea
The magnetic gauge boson
plasma mass $\dt m_{W,T}$ of order
$g^2$ is a non--perturbative feature
of the theory which
was first predicted in \cite{GPY} and
derived in e.g. \cite{BFHW,EQZ}.
The factor $\gm$ is expected to be
of order one as confirmed
by lattice simulations and other
non--perturbative methods \cite{BLS}.

Though fermionic contributions
to the wave function correction term
are not expected to introduce
new infrared singularities,
the top quark self--mass
$\Si_t$ is evaluated in appendix
\ref{topselfen}. We do not
find any plasma mass correction to first
loop--order. This is however
not surprising, since at least in the
symmetric phase $m_t =0$ is
protected to any finite order in
perturbation theory by chiral
invariance. Instead we find a chemical
potential which is
non--vanishing in both phases of the theory. We
will however neglect this
one--loop correction to the tree--level
propagator. In the next
section we will find, that the
(unimproved) top quark
contribution to $Z(\vph,T)$ is indeed
infrared--finite and furthermore
negligible compared to other
contributions. Therefore there
is no need to improve perturbation
theory by including this plasma
induced effective chemical potential
into the top quark propagator.

The improved propagators
for the bosonic degrees of freedom
are now given in
terms of the full masses \be
 m_i^2=m_i^{(0)\, 2}+\dt m_i^2 \; .
\ee
The tree--level
connection between the couplings and
zero--temperature masses is given by
$g=2m_W(T=0)/v$, $f_t=\sqrt{2}m_t(T=0)/v$ and
$\la=m_H^2(T=0)/2v^2$ where $v=\sqrt{-\mu^2
/\la}\simeq 246\GeV$.

\section{Local momentum expansion}
\label{LocMomentExp}

Previous work on terms
quadratic in spatial derivatives in
the effective action include
\cite{MTW} and \cite{CW}--\cite{Zuk}.
We will follow the approach
of Moss et.~al. which is based on the
local momentum space method of \cite{BP}.

We start with a short review
of the methods developed in
\cite{MTW} and generalize them
by including plasma masses.
The calculation of the one--loop
contribution to the wave function
correction term in the effective
action only requires the part
of (\ref{SMLagrangian}) which is bilinear
in the quantum
fluctuations\footnote{Of course,
$h$ and $\chi_i$
are exactly the fluctuations
which are collectively denoted by
$\hat{\vph}$ in (\ref{PartFunct0}).}
$h$, $\chi_i$, $W$, $c$ and
$t$:
\be
 \Lc_2 = \ds{\hal \dt_{ab}W^{a}_{\mu}
 \Dt_{W}^{\mu\nu}W^{b}_{\nu}
 +\dt_{ab}\ol{c}^a \Dt_c c^b
 +\hal h \Dt_\varphi h
 +\hal\sum_j \chi_j \Dt_\chi\chi^j
 +\ol{t}\Dt_t t}
 \label{BilinearAction}
\ee
with
\bea
 \Dt_{W}^{\mu\nu}(x)
 &=& \ds{
 \left(\prl^\rho\prl_\rho
 +m_W^2\right)g^{\mu\nu}
 -\left(1-\frac{1}{\xi}\right)
 \prl^\mu\prl^\nu}\nnn
 && \ds{-\frac{2}{\xi}
 \left( N^\mu\prl^\nu-N^\nu\prl^\mu\right)
 +\frac{2}{\xi}\left(\prl^\nu N^\mu\right)
 -\frac{4}{\xi}N^\mu N^\nu}\nnn
 \Dt_c(x) &=& -\ds{\left(
 \prl^\rho\prl_\rho+\xi m_W^2
 +2N^\mu\prl_\mu\right)}\nnn
 \Dt_{\vph,\chi}(x) &=&
 \ds{-\prl_\mu\prl^\mu -m^2_{\vph,\chi}}\nnn
 \Dt_t(x) &=& \ds{i\slash{\prl}+m_t}
 \label{InvPropag}
\eea
and $N_\mu=\prl_\mu\ln\vph$.
The one--loop corrections to
the effective action may then be
written as
\be
 \Gm^{(1)}=-\sum_n a_n\tr\left(\ln\Dt_n(x)\right)
 \label{Gamma}
\ee
where $n$ runs over all fields in the theory.
As seen above, $\Dt_n$ is
an operator of the general form
$\Dt_n(x)=D_n+M_n(x)$ with $D_n$ being
a differential operator and
$M_n(x)$ a mass term\footnote{Note that
$M_n(x)$ actually corresponds
to a mass term only for fermions but
to a mass {\em squared} term for
gauge bosons and scalars.}. The
coefficients $a_n$ are fixed by
the statistics and the number of
degrees of freedom of the
corresponding field. In particular, we
have $a_\vph=a_{\chi_j}=\hal$,
$a_W=\frac{3}{2}$ and $a_t=-1$.

The effects of the plasma
are taken into account by
using the inverse
propagators\footnote{One may equivalently use
improved (inverse) propagators
from the beginning in
(\ref{BilinearAction}).
In this case one has to add appropriate
plasma mass counter terms
to (\ref{BilinearAction}) in order to
compensate for these corrections \cite{BFHW}.}
\be
 \Dt_n =\left( D_n+M_n+\Si_n\right) -\Si_n \equiv
 \Dt^{\prime}_n -\Si_n(x) \; ,
\ee
where $\Si_n$ denotes all
plasma corrections for the inverse
propagator, and may in general depend on $x$ via
$\vph$. Therefore, to
one--loop, (\ref{Gamma}) is equivalent to the
modified relation
\be
 \Gm_n^{(1)}=-a_n
 \tr\left[\left( 1
 -\Si_n\frac{\prl}{\prl\Si_n}\right)
 \ln\Dt_{n}^{\prime}\right] \; .
 \label{1loopEffAction}
\ee

The situation for
gauge fields is slightly more
complicated due to the existence of
different plasma masses for
transverse and longitudinal degrees of
freedom. According to
(\ref{GaugeProp}), (\ref{WSelfEn}) the
operator
\be
 \Dt_W=\left(\Dt_{W,L}^{\prime}
 -\dt m_{W,L}^2 P_L\right)
 +\left(\Dt_{W,T}^{\prime}
 -\dt m_{W,T}^2 P_T\right)
 +\Dt_{W,G} +\ldots\; ,
\ee
where the dots indicate terms
depending on $N_\mu$, splits up into
three contributions:
\bea
 \Dt_{W,L}^{\prime} &=& \ds{
 \left(\prl^\rho\prl_\rho+m^2_{W,L}\right) P_L}\nnn
 \Dt_{W,T}^{\prime} &=& \ds{
 \left(\prl^\rho\prl_\rho+m^2_{W,T}\right) P_T}\nnn
 \Dt_{G} &=& \ds{\frac{1}{\xi}
 \left(\prl^\rho\prl_\rho
 +\xi m^{(0)2}_{W}\right) P_G}\; .
\eea
In Landau gauge the final
result does not depend on the last term
and thus we need no counter terms for $\Dt_{G}$.

The first step in evaluating
(\ref{1loopEffAction}) is the definition of
a (finite--temperature) Greens function
$G_n(x,\pr{x})$ for each $\Dt^{\prime}_n$ via
\be
 \Dt^{\prime}_n G_n(x,\pr{x})=
 \delta_{x_0,\pr{x}_0}\delta (\vec{x}-\pr{\vec{x}})\; .
 \label{GreensFunction}
\ee
If we furthermore assume that
$M_n$ may be written as
$M_n(x)=m_n+\ol{M}_n(x)$,
where $m_n$ is independent of
$x$ and $\ol{M}_n$ is independent
of $m_n$, we arrive at
\be
 \frac{\prl \Gm^{(1)}_n}{\prl m_n}=
 -a_n\int_\beta
 \tr\left[\left( 1-\Si_n\frac{\prl}
 {\prl\Si_n}\right)
 G_n(x,x)\right] \; ,
 \label{Effective Action}
\ee
where the functional trace is
accounted for by the
integral. Hence the knowledge
of the $G_n(x,\pr{x})$ will enable us
to derive the one--loop contribution
$\Gm^{(1)}_n$ to the effective
action via
integration w.r.t. $m_n$.

Since we only need to know the\footnote{For
simplicity we will suppress
the index $n$ from now on. It is clear
that the following steps have
to be carried out for each degree of
freedom labeled by $n$, separately.}
$G(x,\pr{x})$ in the limit
$x\ra\pr{x}$, we expand the functions
$\pr{M}(x):=M(x)+\Si(x)$ into a
Taylor series around
$x=\pr{x}$:  \be
 \pr{M}(x)=\pr{M}(\pr{x})+\sum_{l=1}^\infty
 \frac{1}{l!}\pr{M}_{\mu_1\ldots\mu_l}(\pr{x})
 y^{\mu_1}\ldots y^{\mu_l}
\ee
with $y=x-\pr{x}$ and
\be
 \pr{M}_{\mu_1\ldots\mu_l}(\pr{x})=\left.
 \frac{\prl^l \pr{M}(x)}
 {\prl x^{\mu_1}\ldots\prl x^{\mu_l}}
 \right|_{x=\pr{x}} \; .
\ee
Inserting this into (\ref{GreensFunction})
and Fourier transforming
w.r.t.~$x$ (denoted by a tilde)
\be
 G(x,\pr{x})=\sum_{k_0}\frac{1}{\beta}
 \int\frac{d^3 k}{(2\pi)^3}
 e^{i(k_0 y_0-\vec{k}\vec{y})}
 \tilde{G}(k,\pr{x}) \; ,
 \label{FourierTrans}
\ee
we arrive at
\be
 \left( \tilde{D}+\pr{M}(\pr{x})+
 \sum_{l=1}^\infty \frac{i^l}{l!}
 \pr{M}_{\mu_1\ldots\mu_l}(\pr{x})\frac{\prl^l}
 {\prl k_{\mu_1}\ldots\prl k_{\mu_l}}\right)
 \tilde{G}(k,\pr{x})=1 \; .
\ee
Note that the sum in (\ref{FourierTrans})
may be over bosonic
($k_0=2\pi inT$) or fermionic
($k_0=(2n+1)\pi iT$) Matsubara
frequencies, depending on the
type of inverse propagator
$\Dt^{\prime}$ used to define
$G(x,\pr{x})$. We may now expand
$\tilde{G}(k,\pr{x})$ and $\pr{M}(\pr{x})$
in a series w.r.t.~the
number of derivatives\footnote{For
scalar and fermion fields the
derivative expansion of $\pr{M}$
is trivial. Eqn. (\ref{InvPropag})
however shows that it is
non--trivial for vector and ghost fields.}:
\bea
 \pr{M}(x^{\prime}) &=& \ds{\sum_{l=0}^\infty
 i^l M^{\prime (l)}(x^{\prime})}\nnn
 \tilde{G}(k,\pr{x}) &=& \ds{\sum_{j=0}^\infty
 \tilde{G}^{(j)}(k,\pr{x})} \; .
 \label{DerivativeExpansion}
\eea
One immediately obtains
\be
 \tilde{G}^{(0)}(k,\pr{x})=
 \frac{1}{\tilde{D}(k)+\pr{M}(\pr{x})}
\ee
and
\be
 \tilde{G}^{(j)}(k,\pr{x})
 =-\tilde{G}^{(0)}(k,x^{\prime})
 \sum_{s=1}^j \sum_{l=0}^s\frac{i^s}{l!}
 M_{\mu_1\ldots\mu_l}^{\prime (s-l)}
 (\pr{x})\frac{\prl^l}
 {\prl k_{\mu_1}\ldots\prl k_{\mu_l}}
 \tilde{G}^{(j-s)}(k,\pr{x})\; .
 \label{IterativeSol}
\ee
This iterative solution of
(\ref{GreensFunction}) may now be used
to determine the Greens functions
and subsequently the
effective action via
(\ref{Effective Action}) order by order in the
derivative expansion (\ref{DerivativeExpansion}).
The $\tilde{G}^{(0)}(k,\pr{x})$
yield the effective potential whereas the
$\tilde{G}^{(2)}(k,\pr{x})$ will give the
desired kinetic term in the
one--loop effective action:
\be
 \Gm^{(1)}_{\rm kin} = \ds{-\sum_n a_n
 \int dm\int_\beta}\;
 \ds{
 \lim_{y\ra 0}\sum_{k_0}\frac{1}{\beta}
 \int\frac{d^3 k}{(2\pi)^3}
 e^{i(k_0 y_0-\vec{k}\vec{y})}
 \tr\left[\left( 1-\Si_n\frac{\prl}{\prl\Si_n}\right)
 \tilde{G}_n^{(2)}(k,\pr{x})\right]} \; .
 \label{FinalResult}
\ee

Although (\ref{IterativeSol}) and
(\ref{FinalResult}) provide a
well--defined prescription
for calculating the wave function
correction, it still requires
rather lengthy calculations. We have
used the symbolic manipulation
package {\em Mathematica} to
accelerate the computation.
All that is needed are the (improved)
zeroth order propagators
$\tilde{G}_n^{(0)}(k,x)$ as given in
(\ref{ImprovedScalarProp}),
(\ref{GaugeProp}) as well as the
standard tree level top quark propagator, and
the space--time dependent pieces
$M_n^{\prime (l)}(x)$ which may be
read off directly from the results of section
\ref{SelfEnergyCorrections}. As
explained in section \ref{Inhom} we
will only take into account stationary fields,
i.e., $\prl_0\vph=0$,
and use standard
integration techniques for the
finite--temperature momentum
integrals. All contributions split up
into a $T$--independent and a
$T$--dependent piece as is generally
the case in thermal field theory.

The temperature dependent scalar contribution turns
out to be
\be
 Z_{\rm scalar}^{T\neq 0}(\vph,T)=
 \frac{\la^2 T\vph^2}{16 \pi}
 \left(
 \frac{9\dt m_\varphi^2}{2m_\varphi^5}
 +\frac{3}{m_\varphi^3}
 +\frac{3\dt m_\chi^2}{2m_\chi^5}
 +\frac{1}{m_\chi^3}\right) \; .
\ee
The corresponding $T$--independent
part is UV--finite and given by
\be
 Z_{\rm scalar}^{T=0}(\vph)=
 \frac{\la^2\vph^2}{16\pi^2}
 \left(\frac{3}{m_\vph^2}
 +\frac{3\dt m_\vph^2}{m_\vph^4}
 +\frac{1}{m_\chi^2}+\frac{\dt m_\chi^2}{m_\chi^4}
 \right) \; .
\ee
Since it is subleading in the high--$T$ expansion
compared to $Z_{\rm scalar}^{T\neq 0}
(\vph,T)$ we will neglect it
in the following. The $T$--dependent
gauge boson one--loop
contribution turns out to be
\be
 Z_{\rm vector}^{T\neq 0}(\vph,T) =
 -\frac{3g^2T}{4\pi}
 \left(
 \frac{m_W^{(0)4}}{32m_L^5}
 -\frac{5m_W^{(0)2}}{96m_L^3}
 +\frac{5m_W^{(0)4}}{16m_T^5}
 -\frac{41m_W^{(0)2}}{48m_T^3}
 +\frac{1}{m_T}\right) \; .
 \label{ZvectorTneq0}
\ee
The zero--temperature contribution
is UV--divergent and has to be
renormalized. It can be seen
to be of the form
\cite{MTW}
\be
 Z_{\rm vector}^{T=0}(\vph)\sim
 \ln\left(\frac{m_W}{m_{\rm ren}}\right)\; ,
\ee
where $m_{\rm ren}$ denotes the renormalization
scale. Compared to
(\ref{ZvectorTneq0}) this part of
the wave function correction term
is again subleading in the high--$T$
expansion and may therefore be
neglected.

We now turn to the one--loop
top quark contribution
$Z_{\rm fermion} (\vph,T)$. In the
high--temperature expansion
its leading order $T$--dependent
piece is given by:
\be
 Z_{\rm fermion}^{T\neq 0}(\vph,T)=
 \frac{f_t^2}{24\pi^2}
 \left( 1+3\gm_E+3\ln\frac{m_t}{\pi T}
 \right)\; ,
 \label{TopZfactor}
\ee
which is logarithmically
IR--divergent. However, in
this case the renormalized--temperature
contribution which
could be neglected for the other
fields comes to the rescue. It is
given by
\be
 Z_{\rm fermion}^{T=0}(\vph)=
 \frac{f_t^2}{32\pi^2}\left(
 1-4\ln\frac{m_t}{m_{\rm ren}}\right)\; ,
\ee
and therefore cancels exactly
the IR--divergence of the
$T$--dependent part. The full
fermionic contribution is of the form
\be
 Z_{\rm fermion}(\vph,T)=\frac{f_t^2}{96\pi^2}
 \left( 7+12\gm_E+12\ln\frac{m_{\rm ren}}
 {\pi T}\right)\; .
\ee

We would like to point out
that our results for the bosonic as well
as the $T$--independent
fermionic contributions to $Z(\vph,T)$
agree with the findings of
\cite{MTW} for vanishing plasma masses.
However, the fermionic
$T$--dependent piece (\ref{TopZfactor})
disagrees with that of
\cite{MTW} where a $1/m_t$
infrared singularity (which can not be
canceled by the $T$--independent
part) was found. The reason for
this discrepancy appears to be
that in \cite{MTW} the kinetic
energy was calculated following
(\ref{FinalResult}) by summing over
{\em bosonic} frequencies $k_0$
instead of {\em fermionic}
frequencies as required by the
Fourier transformation
(\ref{FourierTrans}) of a
fermionic Greens function.
Our results
show that all infrared singularities
in the symmetric phase are
removed by the plasma masses to
leading order in the high--$T$ and
the loop--expansion.

The full wave function correction
term is now given by
\be
 Z(\vph,T)=Z_{\rm vector}(\vph,T)
 +Z_{\rm fermion}(\vph,T)\; .
 \label{Zfull}
\ee
Note that here $Z_{\rm scalar}$ is not
included.
We wish to emphasize
that it would be inconsistent to
add the scalar contribution for
the calculation of physical
quantities like the free energy, the
droplet tension or droplet
nucleation rate, following the methods
described in section
\ref{Inhom}.\ref{DecayMetaStates}. The scalar
contributions to the effective
action are fully taken into account
by solving the bosonic path
integral (\ref{PartFunct}) in the
saddle point approximation.
However, for checking the range of
validity of the derivative
expansion  (\ref{DerExpansion}) one
might as well integrate out
the scalars in the same way as the
other fields and consider the quantity
\be
 \tilde{Z}(\vph,T)=Z_{\rm scalar}(\vph,T)
 +Z_{\rm vector}(\vph,T)+Z_{\rm fermion}(\vph,T)\; .
\ee

\section{Results}
\label{Results}

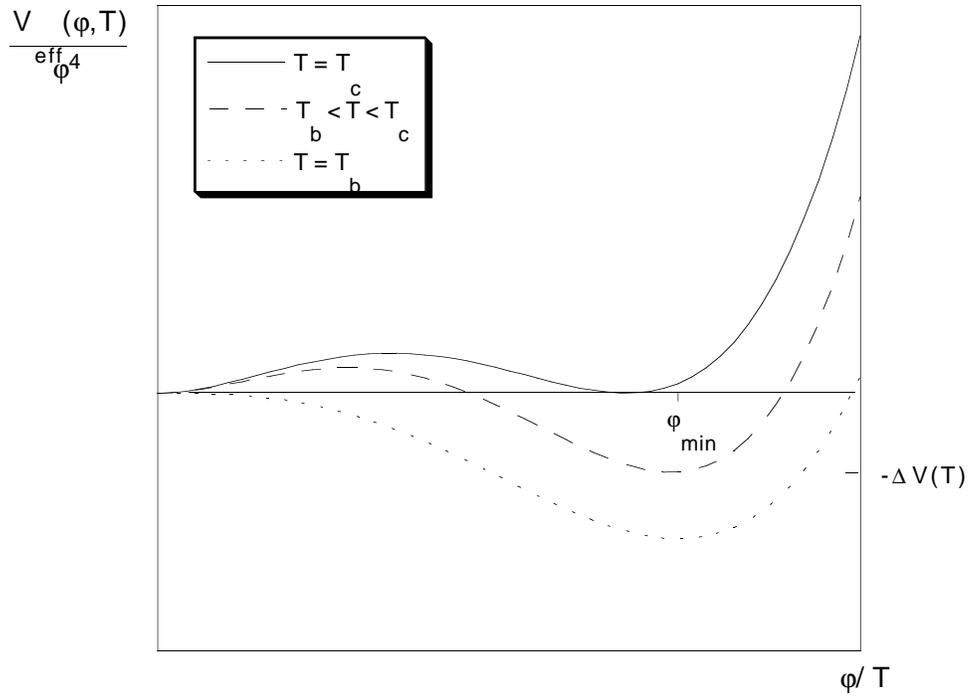
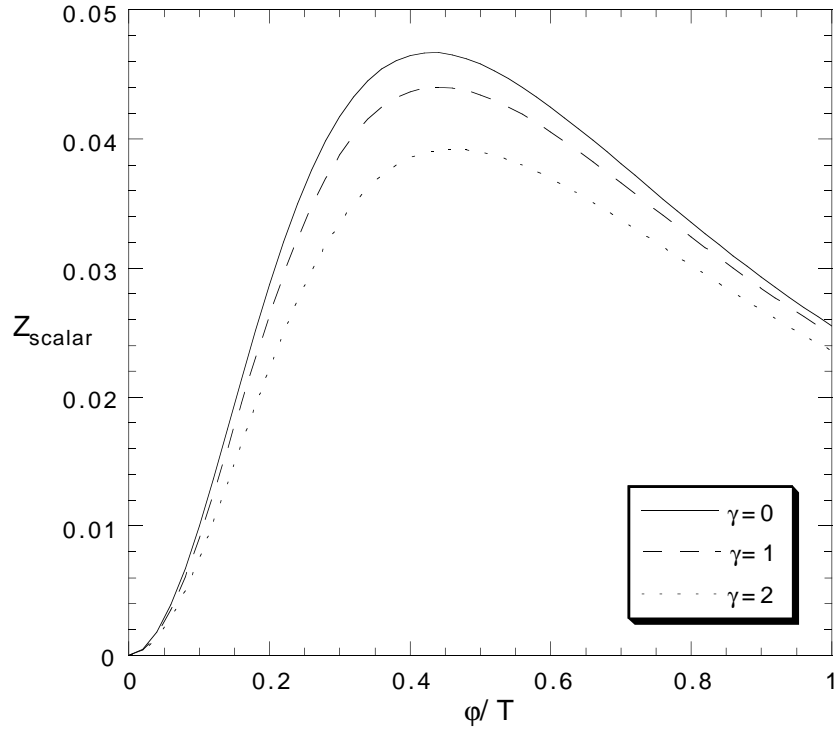
\begin{figure}
 \caption{\em The total wave function
 correction factor $Z(\vph,T)$
(2.a), its scalar (2.b) and vector
(2.c) field contribution
are plotted for
$m_H(T=0)=85\GeV$, $m_t(T=0)=160\GeV$
and $\gm=0,1,2$ at the
corresponding critical temperature
$T=T_c$. Note that $T_c$
is a function of $\gm$, so that also
$Z_{\rm scalar}(\vph,T_c)$ is
implicitly $\gm$--dependent.}
\label{Zfactor}  \end{figure}

A short look at the various
contributions to $Z(\vph,T)$ shows that
it is completely dominated by
its gauge boson contribution $Z_{\rm
vector}(\vph,T)$ for
temperatures close to the critical temperature
$T_c$. This is demonstrated
in Fig.~\ref{Zfactor}, where $Z$ and
all its contributions are
plotted for physically realistic values
of the parameters $m_H(T=0)$,
$m_t(T=0)$ and $\gm$ at $T=T_c$. We
also note that an inclusion of
$Z_{\rm scalar}$ would be numerically
insignificant.

Since the perturbative expansion of the wave
function correction term starts
at the one--loop level, the size of
$Z(\vph,T)$ is a direct measure
for the convergence of perturbation
theory. Furthermore, for values
of $\vph$ and $T$ for which
$\abs{Z(\vph,T)}\gta 1$ or
$\abs{\tilde{Z}(\vph,T)}\gta 1$,
we have to expect that other
higher derivative terms of the
effective action in
(\ref{DerExpansion}) could also
become important and may no longer
be neglected. It is already known
\cite{BFHW}, that there is only a
small window of parameters close
to the current experimental bound
on the Higgs mass, where
perturbation theory can be trusted as far
as a calculation of the
effective potential is concerned. In
particular, for Higgs masses
above approximately $85\GeV$ higher
order loop--contributions to
the effective potential become dominant
and perturbative statements
on the phase transition are
questionable. A first order
phase transition can reliably be found
for $m_H\lta 85\GeV$ and small
$\gm$. It is altogether excluded
for $\gm\gta 2$ taking into
account the current lower bound on the
Higgs mass of $m_H\gta 63.5\GeV$ \cite{LEPHiggs}.

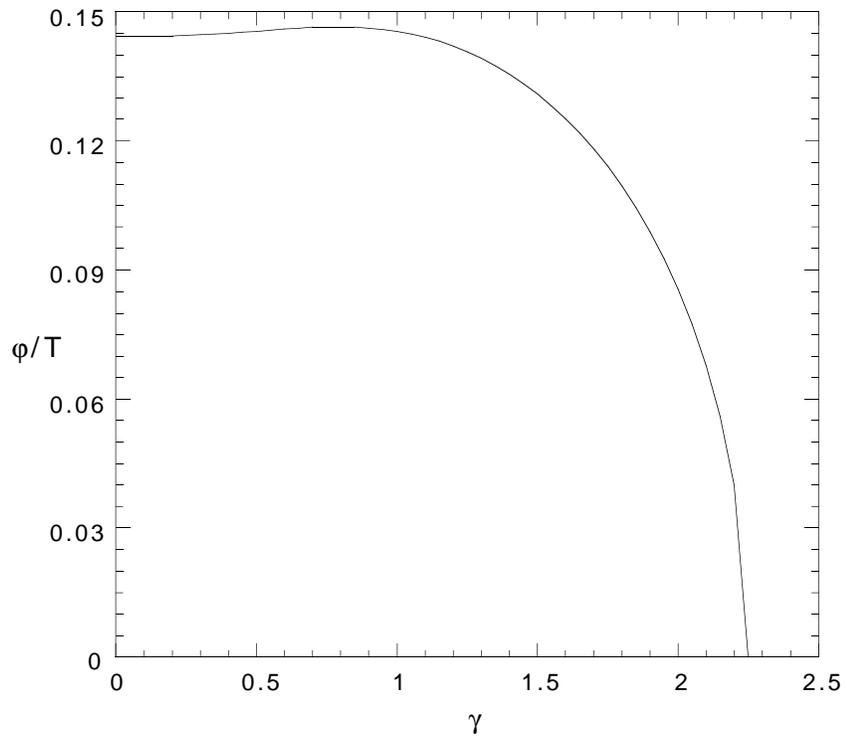
\begin{figure}
 \caption{\em The plotted
 boundary curve divides $\gm$--$\vph$
space into two regions. Above the boundary
$\abs{Z(\vph,T)}\simeq
\abs{Z_{\rm vector}(\vph,T)}<1$. Below it
$\abs{Z(\vph,T)}>1$ and the wave
function correction term
dominates over the classical
kinetic term of the effective action.}
\label{Bound}
\end{figure}

In Fig. \ref{Bound} we have
plotted the curve which divides
$\gm$--$\vph$--space into two
regions. Above it
$\abs{Z(\vph,T)}<1$ and the
wave function correction term
is smaller than the classical kinetic
term of the effective action.
Below this curve $\abs{Z(\vph,T)}>1$,
perturbation theory breaks down
and higher derivative terms might
dominate the dynamics of the
phase transition. We have used
$\abs{Z(\vph,T)}\simeq\abs{Z_{\rm vector}
(\vph,T)}$, which is a
good  approximation for all
temperatures relevant for the phase
transition. We find that for
the whole range of magnetic masses
which permit a first--order
phase transition (taking into account
the experimental lower bound
on the Higgs mass), there is an
interval of small $\vph$--values
for which the wave function
correction dominates over the
classical kinetic term and is
furthermore negative. Therefore,
in this range the field
transformation (\ref{ScalarTrans})
yields an imaginary scalar field
$\tilde{\vph}$ and a complex
effective potential
$\tilde{V}(\tilde{\vph},T)$.
This in turn renders the methods of
section \ref{Inhom} inapplicable.
However, a rough estimate, using
an averaged (constant) wave
function correction factor
$<Z(\vph,T)>=\frac{1}{\vph_{\rm min}}
\int_0^{\vph_{\rm min}}d\vph\,
Z(\vph,T)$, gives e.g. for $\gm=1$,
$m_H=85\GeV$ and $m_t=160\GeV$
the value $<Z(\vph,T)>\approx -0.4$.
This in turn yields an
uncertainty of the derivative
expansion of approximately $50\%$
(c.f. (\ref{Criterion})) and a
weakening of the phase
transition strength
$\frac{\vph(T_e)}{T_e}$ by $3\%$.

At the one--loop level, we
are thus left with the following
situation: For small magnetic
masses ($\gm\lta 2$), i.e., for the
range of $\gm$ which, in
combination with the current experimental
lower bound on the Higgs--mass,
allows for a first--order phase
transition, neither perturbation
theory nor the derivative
expansion seem to be reliable for
small scalar field values.
In particular, for those values of
$\vph$ which are important for
the determination of the dynamics
of the phase transition,
$0\leq\vph\leq\vph_{\rm min}$,
the one--loop wave function
correction term dominates over
the classical kinetic term
in the effective action.
Therefore, in this range of parameter
space, it is not possible to
make physically reliable predictions at
the one--loop level. In order
to clarify the dynamics of a
first--order phase transition,
clearly, the calculation of higher
loop--corrections to $Z(\vph,T)$
as well as higher derivative terms
of the effective action are necessary.
However, since the
one--loop result for $Z(\vph,T)$
is negative for the whole range of
$0\leq\vph/T\leq 1$, we expect that
its inclusion will {\em weaken}
the first--order electroweak
phase transition, as explained in
section \ref{Inhom}.

For larger values of the
magnetic mass, $\gm\gta 2$, the
perturbative calculation of
the wave function correction term
appears to be reliable. On the
other hand such magnetic masses
correspond to a part of
parameter space, where the phase transition
is expected to be of second
order \cite{BFHW}, and baryo--genesis
can not appear at the electroweak
scale within the mechanism of
\cite{KRS}, \cite{FS}.

\section{Conclusions}
\label{Conclusions}

We have investigated the
impact of the scalar field wave
function correction term $Z(\vph,T)$ on the
electroweak phase transition.
Since $Z(\vph,T)$ is the quantum
correction to the classical
scalar field kinetic term in the
effective action, its knowledge
is crucial for an understanding of
the influence of inhomogeneous
scalar field configurations on the
dynamics of this transition.

It turned out that generally a positive
$Z$--factor would increase
the strength of a possible first--order
transition whereas a negative
$Z$ would decrease it. A calculation
of $Z(\vph,T)$ within the
Standard Model to one--loop revealed that
it is fully dominated by
its gauge boson contribution for
temperatures close to the
critical temperature $T_c$ and well above
the barrier temperature $T_b$.
Furthermore we find that
the one--loop result for
$Z(\vph,T)$ is negative for the whole range
$0\leq\vph /T\leq 1$ and
$T\simeq T_c$. However, for small magnetic
gauge boson masses, $\gm\lta 2$,
which would allow for a first--order
phase transition for appropriate Higgs
masses, we find
$\abs{Z(\vph,T)}\gta 1$ for
small $\vph$. We interpret this fact as
an indication that higher
loop--corrections as well as higher
derivative terms in the
effective action might be crucial for an
understanding of the dynamics
of the phase transition. For larger
magnetic masses we obtain
$\abs{Z(\vph,T)}\lta 1$, and conclude that
in this part of parameter
space perturbation theory seems to be
reliable. However, for such
values of the magnetic gauge boson
masses the phase transition
is presumably of second order.

Our negative one--loop result
for $Z(\vph,T)$ for the
range $0\leq\gm\lta 2$ seems to
indicate a decrease of the strength
of the first--order phase transition.
A further clarification of
this point would however
require the knowledge of higher
loop--corrections to $Z(\vph,T)$
and perhaps also higher derivative
terms in the effective action.

\vspace{.5 cm}
\ \\
\bf{Acknowledgment}: \normalsize
We acknowledge valuable discussions
with Scott Dodelson and Eric Weinberg.

\newpage

\section*{Appendix}

\appendix

\section{The top quark self--mass}
\label{topselfen}

In this appendix we evaluate
the top quark self--mass
$\Si_t$. Among the various
one--loop Feynman diagrams contributing
to $\Si_t$ the dominant contribution
comes from the gluon
loops\footnote{For a very heavy
top quark with $m_t\simeq 200\GeV$
one expects the Higgs loop to
give a comparable contribution, since
then $f_t\simeq g_s$.}.
The corresponding contribution to $\Si_t$
has been calculated in \cite{Weldon}
for the symmetric phase. For
$SU_c(3)$ one obtains
\be
 \Si_t^{\rm gluon}(k_0,0)=\frac{1}{6}
 \frac{g_s^2 T^2}{k_0} \gm_0\; .
 \label{SU3SelfMass}
\ee
Hence the gluon loops do not
induce a plasma mass but an
effective chemical potential
$\mu(k_0,\vec{k}=0)=\frac{1}{6}
\frac{g_s^2 T^2}{k_0}$. The
absence of a plasma mass in the
symmetric phase is not surprising,
because we do not expect chiral
invariance to be broken to any
finite order in perturbation
theory. For the same reason the
$SU_L(2)$ gauge boson
loop--corrections to $\Si_t$ should not
induce a plasma mass.
Since furthermore $g<<g_s$, they will be
neglected in the following,
and we will concentrate on the scalar
loop--corrections to $\Si_t$.

The Higgs boson loop--contribution is given by
\be
 \Si_t^{\rm Higgs} (k_0,\vec{k}) =
 \frac{1}{4}f_t^2 \sum_{p_0} T
 \int\frac{d^3 p}{(2\pi )^3}
 \frac{-i(\slash{p}+m_t)}{p^2-m_t^2}
 \frac{i}{(p-k)^2-m_\vph^2}\; ,
\ee
where $p_0=(2n+1)i\pi T$ and
$k_0=(2m+1)i\pi T$ are fermionic
Matsubara frequencies.
The determination of a plasma
mass or chemical potential
requires the knowledge of $\Si_t(k_0,0)$:
\be
 \Si_t^{\rm Higgs} (k_0,0)=\frac{1}{4}f_t^2 \left[
 m_t+\frac{\gamma_0}{2k_0}
 \left( m_t^2-m_\vph^2+k_0^2\right)\right] I_1 -
 \frac{1}{8k_0}f_t^2\gamma_0  I_2
\ee
with
\bea
 I_1&=&\ds{\sum_{n=0}^\infty T
 \int\frac{d^3 p}{(2\pi )^3}
 \frac{1}{(p^{(-)2} -m_t^2)(p^{(+)2} -
 m_\vph^2)}} \nnn
 I_2&=&\ds{\sum_{n=0}^\infty T
 \int\frac{d^3 p}{(2\pi )^3}
 \left(\frac{1}{p^{(-)2} -m_t^2}-
 \frac{1}{p^{(+)2} -m_\vph^2}\right)} \; .
 \label{integrals}
\eea
Fermionic four--momenta are denoted by
$p^{(-)}=(p_0^{(-)},\vec{p})$
whereas bosonic momenta are denoted as
$p^{(+)}=(p_0^{(+)},\vec{p})$
with Matsubara frequencies
$p_0^{(-)}=(2n+1)i\pi T$, $p_0^{(+)}=2ni\pi T$.
The evaluation of $I_2$ in
the high--temperature expansion is
straightforward, whereas $I_1$
is somewhat more involved and given
in appendix \ref{someintegrals}. Using these
results we obtain to leading
order in $f_t$, $y_b=m_\vph /T$ and
$y_f=m_t/T$
\be
 \Si_t^{\rm Higgs} (k_0,0) = \frac{f_t^2}{32\pi}Cy_f
 T + \frac{f_t^2}{32\pi}\gamma_0 \left[
 \hal C k_0-k_0\frac{y_b}{\pi}
 -\frac{T^2}{k_0}\left(\hal
 +\frac{y_b}{\pi}\right)\right]
 \label{TopSelfEnergy}
\ee
with $C=-1+\gamma_E+\ln 2$.

The physical degrees of freedom are
found by determining the poles of the
propagator
$S_F (k_0,\vec{k}) =[\slash{k}-m_t
-\Si_t (k_0,\vec{k})]^{-1}$
for $\vec{k}=0$.
A straightforward calculation reveals
four different zeros of $S_F^{-1}$.
In the symmetric phase, i.e.,
for $y_f=y_b=0$, the pole
positions are degenerate and
given by $k_0=\pm f_t T/8$,
which is in perfect agreement with the
findings of
\cite{Weldon,BBS}\footnote{Actually the degeneracy
already occurs for $y_f=0$
but $y_b\neq 0$. Furthermore it is
lifted even in the symmetric
phase for $\vec{k}\neq 0$
\cite{Weldon}.}.

The form of $\Si_t(k_0,0)$ shows an
additional plasma correction
to the effective chemical
potential as well as a plasma
mass for the top quark of the
form
\be
 \mu_t = \frac{f_t^2}{32\pi}\left(\hal
 -\frac{y_b}{\pi}\right)\frac{T^2}{k_0}
 -\frac{f_t^2}{32\pi}\left(\hal
 C-\frac{y_b}{\pi}\right) k_0\; ,\;\;\;
 \dt m_t = \frac{f_t^2}{32\pi}
 C y_f T \; .
\ee
As expected, $\dt m_t$
vanishes in the symmetric phase.
Furthermore one easily sees
that it is exactly canceled by the
neutral Goldstone boson
loop--contribution to $\Si_t$,
which may be calculated
together with the charged
Goldstone contributions
analogously to $\Si_t^{\rm
Higgs}$. The final result is \bea
 \Si_t(k_0,0) = \ds{
 \frac{1}{6}\frac{g_s^2 T^2}{k_0}\gm_0} &+&
 \ds{\frac{f_t^2}{32\pi}\gm_0
 \left[ \;\;C
 k_0-\frac{1}{\pi}y_b k_0-\frac{T^2}{k_0}
 \left(1+\frac{1}{\pi}y_b\right)\right]
 \frac{1+\gm_5}{2}}\nnn
 &+& \ds{\frac{f_t^2}{32\pi}\gm_0
 \left[ 2C
 k_0-\frac{1}{\pi}y_b k_0-\frac{T^2}{k_0}
 \left(2+\frac{1}{\pi}y_b\right)\right]
 \frac{1-\gm_5}{2}}\; .
\eea

\addtocounter{appendix}{1}
\section{Some useful integrals}
\label{someintegrals}

In this appendix we provide the
high--temperature expansion
of the integral $I_1$ defined in
appendix \ref{topselfen} and
for some related integrals also
containing fermionic as
well as bosonic mo\-men\-ta\footnote{Note that
$p_0^{(-)}-k_0^{(-)}$ is bosonic,
though $p_0^{(-)}$ and $k_0^{(-)}$
are fermionic.}. Converting the
Matsubara frequency summation into a
contour integral and reading off
the residues, one finds after
evaluation of the spatial angular integral
\be
 I_1 =\frac{1}{8\pi^2}\left\{
 \left(\Dt y^2-\pi^2\right)
 I_3^{(b)}(y_b,z_b)-\left(\Dt y^2+\pi^2\right)
 I_3^{(f)}(y_f,z_f)\right\}
\ee
with $\Dt y^2=y_b^2-y_f^2$ and
$z_{b/f}^2=\pm\hal\Dt y^2 +\frac{1}{4\pi^2}
 \left(\Dt y^2\right)^2 +y_b^2$ and
\be
 I_n^{(b/f)} (y,z):=\int_0^\infty \frac{x^{n-1}}
 {\left(\frac{\pi^2}{4}+z^2\right)
 \sqrt{x^2+y^2}\left(
 1\mp\exp\sqrt{x^2+y^2}\right) } \; .
\ee
The integrals $I_n^{(b/f)} (y,z)$
may be evaluated in the
high--temperature expansion, i.e.,
for $y,z<<1$, as follows:
In a first step one determines
$I_1^{(b/f)}$. Subsequently
$I_{1+2k}^{(b/f)}$ is calculated
for $y=0$ and positive $k$.
The leading $y$--dependence
for arbitrary $k$ may then be
recovered by applying
\be
 I_n^{(b/f)} (y,z)= -\frac{1}{ny}\frac{\prl}{\prl y}
 I_{n+2}^{(b/f)} (y,z) \;
. \ee
In this way one obtains
\bea
 I_1^{(b)} &=& \ds{\frac{1}{\pi}-\frac{2}{\pi^2}
       \left( 2+\ln\frac{y}{4\pi}\right) -
       \frac{2}{\pi}\left( 1
       -\frac{4}{\pi^2}z^2\right)
       \frac{1}{y}}\nnn
       &+& \ds{\left\{ -\frac{4}{\pi^3}
       -\frac{1}{\pi^2}
       +\frac{4}{\pi^4}
       \left( 7-\gm_C +2\ln\frac{y}{4\pi}
       \right)\right\} z^2}\nnn
 I_1^{(f)} &=& \ds{-\frac{1}{\pi}
       -\frac{2}{\pi^2}\ln\frac{y}{2\pi}+
       \left\{\frac{4}{\pi^3}-\frac{1}{\pi^2}+
       \frac{4}{\pi^4}\left( 1+2\gm_C
       +2\ln\frac{y}{2\pi}
       \right)\right\} z^2 }\nnn
       &+& \ds{\left\{ -\frac{2}{\pi^3}
       +\frac{1}{\pi^2}+
       \frac{1}{\pi^4}\left(-2-8\gm_C
       -4\ln\frac{y}{2\pi}+
       \frac{7}{4}\zeta (3)
       \right)\right\} y^2 }\nnn
 I_3^{(b)} &=& \ds{1-\frac{\pi}{2}-\hal\gm_E+
       \frac{\pi}{4}-\hal\ln 2
       +\frac{2}{\pi}y_b}\nnn
       &+& \ds{\left\{\frac{1}{4}
       +\frac{1}{\pi^2}\left(
       2\gm_C -3\right)\right\} z^2
       +\left\{ -\frac{1}{2\pi}+
       \frac{1}{2\pi^2}
       \left( 2\ln\frac{y}{4\pi}
       +3\right)\right\} y^2}\nnn
 I_3^{(f)} &=& \ds{\frac{\pi}{4}
       -\hal\gm_E-\hal\ln 2}\nnn
       &+& \ds{\left\{\frac{1}{4}
       -\frac{1}{\pi^2}\left(
       2\gm_C+1\right)\right\} z^2 +
       \left\{ \frac{1}{2\pi}
       +\frac{1}{2\pi^2}\left(
       2\ln\frac{y}{2\pi}-1\right)\right\} y^2}\nnn
 I_5^{(b)} &=& \ds{\frac{\pi^2}{8}
       \left(\gm_E +\ln 2-
       \frac{10}{3}+\frac{\pi}{2}\right) }\nnn
       &+& \ds{\hal\left\{
       \frac{\pi}{2}-\frac{\pi^2}{8}
       +\left(\gm_E-\gm_C+\ln 2
       -\hal\right)\right\} z^2
       +\frac{3}{4}\left\{\frac{\pi}{2} +
       \gm_E+\ln 2 -2\right\} y^2}\nnn
 I_5^{(f)} &=& \ds{\frac{\pi^2}{8}
       \left\{\frac{2}{3}+\gm_E
       -\frac{\pi^2}{2}+\ln 2\right\} }\nnn
       &+& \ds{\hal\left\{\hal
       +\gm_E+\gm_C-\frac{\pi}{2}
       -\frac{\pi^2}{8}+\ln 2\right\} z^2
       +\frac{3}{4}\left\{ \gm_E
       -\frac{\pi}{2}+\ln 2\right\} y^2} \; ,
\eea
where $\gm_E\simeq 0.577$
denotes Euler's and $\gm_c\simeq 0.916$
Catalan's constant.

\newpage

\begin{figure}
 \caption{\em The qualitative
 shape of the effective potential is
plotted for three different
temperatures: $T=T_c$, $T_b<T<T_c$ and
$T=T_b$. Here $T_b$ denotes
the barrier temperature for which the
barrier between the two minima
vanishes. For $T_b<T<T_c$
we have also indicated the
position of $\vph_{\rm min}$ and the
potential difference $\Dt V(T)$
between the two minima.}
\label{Veff}
\end{figure}

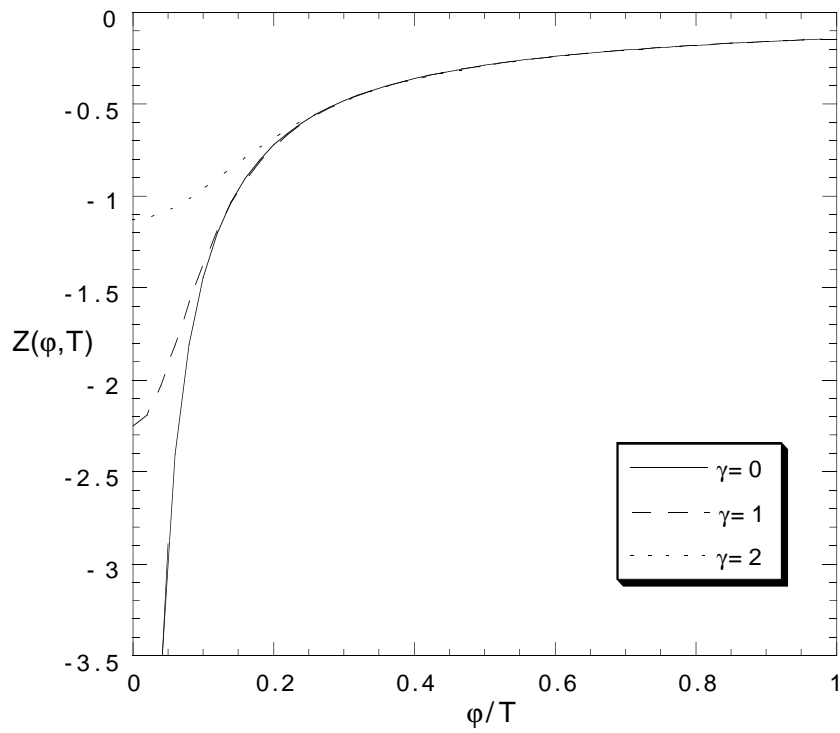
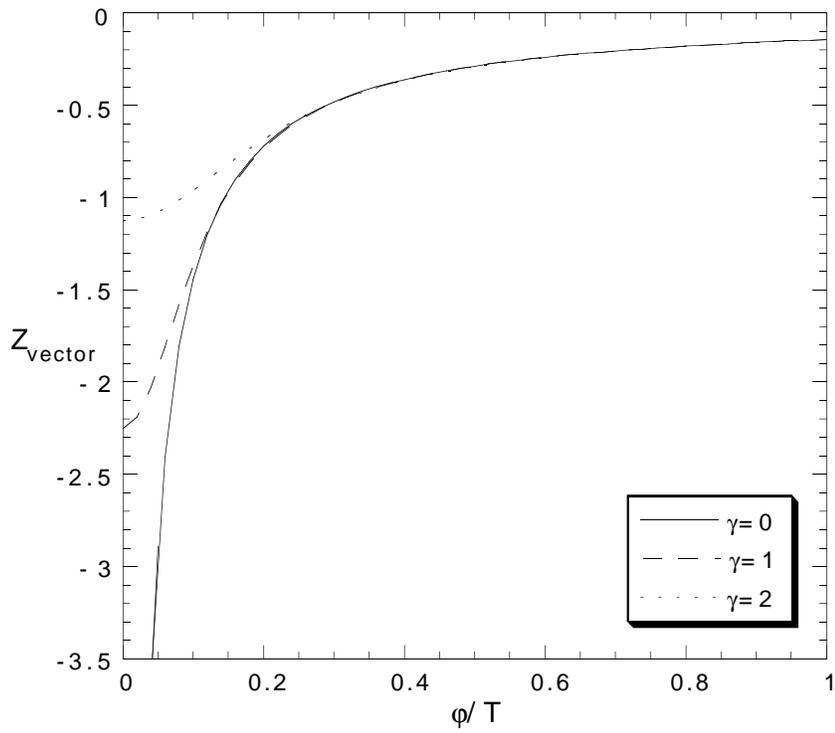
\end{document}